\documentclass[aps,twocolumn,prl,superscriptaddress]{revtex4}
\usepackage{amsfonts}
\usepackage[final,hiresbb]{graphicx}
\usepackage{amsmath}
\usepackage{amssymb}
\usepackage{amstext}

\usepackage{color}
\usepackage{braket}

\definecolor{Green}{RGB}{0,204,102}
\definecolor{Purple}{RGB}{102,0,255}
\definecolor{Blue}{RGB}{51,153,255}
\definecolor{Red}{RGB}{151,010,010}

\begin{document}

\title{Entangled Excitons via Spontaneous Downconversion}

\author{Ariel Shlosberg}
\affiliation{Department of Physics, Colorado School of Mines, Golden, CO 80401, USA}

\author{Mark T. Lusk}
\email{mlusk@mines.edu}
\affiliation{Department of Physics, Colorado School of Mines, Golden, CO 80401, USA}


\begin{abstract}
A class of centrosymmetric molecules support excitons with a well-defined quasi-angular momentum. Cofacial arrangements of these molecules can be engineered so that quantum cutting produces a pair of excitons with angular momenta that are maximally entangled. The Bell state constituents can subsequently travel in opposite directions down molecular chains as ballistic wave packets. This is a direct excitonic analog to the entangled polarization states produced by the spontaneous parametric downconversion of light. As in optical settings, the ability to produce Bell states should enable foundational experiments and technologies based on non-local excitonic quantum correlation. The idea is elucidated with a combination of quantum electrodynamics theory and numerical simulation.
\end{abstract}

\maketitle

The spontaneous parametric down-conversion (SPDC) of light is now routinely used to produce Bell states in which the polarizations of two photons are entangled~\cite{Zeilinger_1995}. This has enabled important experimental studies of the foundations of quantum mechanics, such as the delayed choice quantum eraser~\cite{Scully_1999} and loophole free tests of local realism~\cite{Hensen_2015}, as well as facilitating the development of a range of quantum technologies~\cite{Obrien_2009}. These include quantum network applications such as quantum key distribution and state teleportation~\cite{RevModPhys_74_145}. It has also been shown that orbital angular momentum (OAM) is conserved in SPDC and that the photons produced have entangled OAM states~\cite{Pires_2010}. Such entanglement figures prominently in strategies to increase the number of superimposed states in quantum information processing~\cite{Pires_2010}. 

Recent work has demonstrated that quantum information processing is also feasible using exciton wave packets~\cite{Zang_PRB_2017} with orbital angular momentum transferred back and forth between optical and excitonic manifestations ~\cite{Zang_PRA_2017}.  In particular, excited states of molecules with $C_N$ or $C_{Nh}$ symmetry can be characterized in terms of a quasi-angular momentum. These twisted excitons are phase-shifted superpositions of the lowest energy excitons associated with each of $N$ molecular arms~\cite{AndrewsPRL2013}. Their angular momentum is just the number of $2\pi$ windings of  phase accumulated in traversing the circuit of arms. 

The photonic analogy also carries over to exciton down-conversion in which two low-energy excitons are produced from one of higher energy. This is referred to as multiple exciton generation and carrier multiplication in association with a single structure \cite{Schaller_2004} and quantum cutting (QC) when spatially separated acceptor sites are excited~\cite{Wegh_1999}. However, the focus in such conversions has been on energy splitting and not entanglement. 
Excitonic down-conversion is capable of generating entangled excitonic states though. As in the case of excitonic wave packets, entangled excitonic states may offer ways of manipulating information that is either difficult or impossible in optical settings. This is because electronic states readily couple to a number of easily controlled fields. 

\begin{figure}[hptb]\begin{center}
\includegraphics[width=0.3\textwidth]{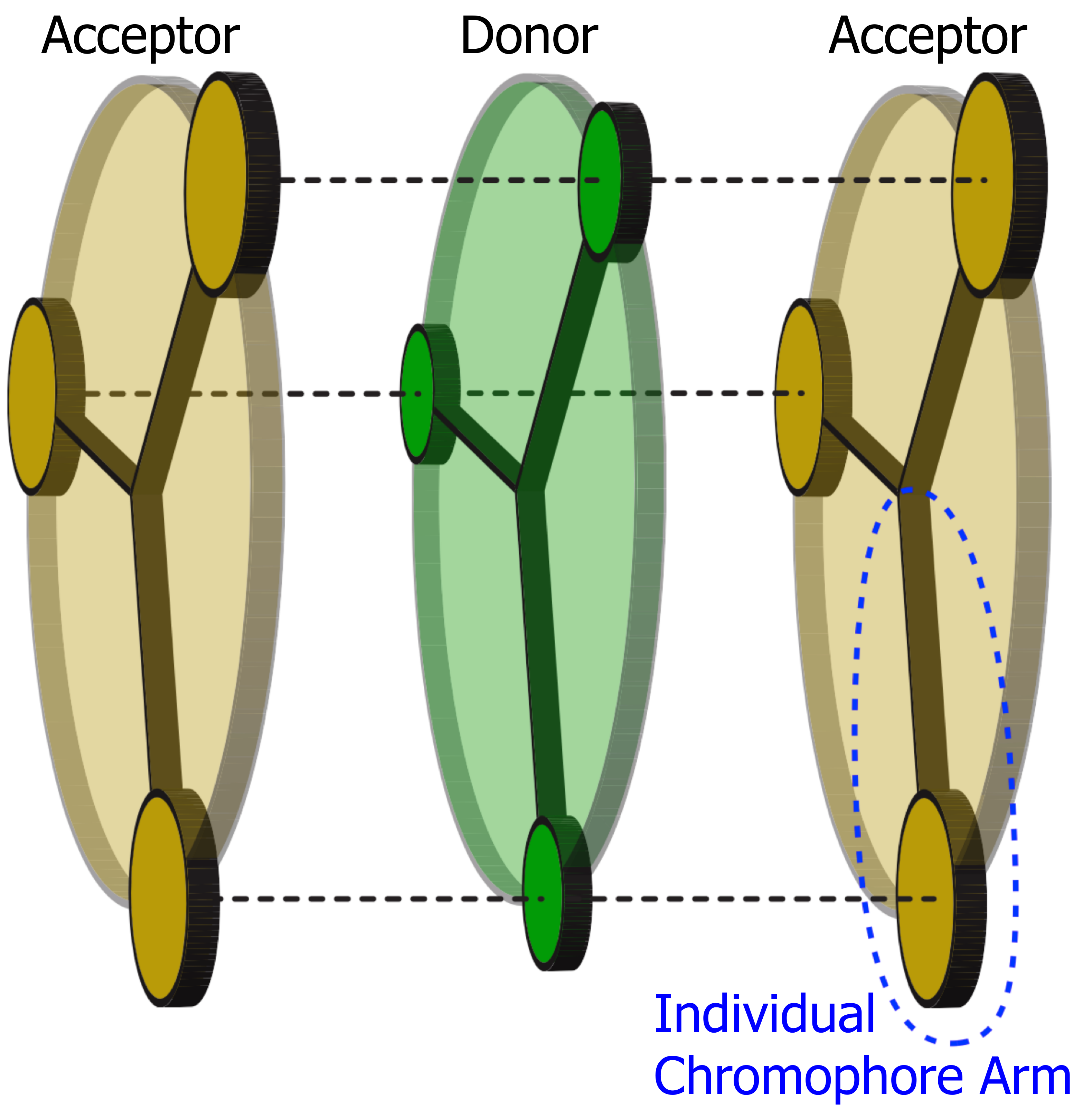}
\caption{\emph{Donor/acceptor system.} Schematic view of three-arm donor molecule sandwiched between acceptor molecules of the same symmetry, $C_3$ or $C_{3h}$.}
\label{Schematic}
\end{center}
\end{figure}

In the present work, we focus on a particularly simple framework to show that quantum cutting on centrosymmetric molecules can be used to generate pairs of excitons with maximally entangled angular momentum. This is accomplished by sandwiching a donor molecule between two acceptors of the same symmetry, as illustrated in Figure \ref{Schematic}. Under the assumption that only the interactions between nearest neighbors matter, excitonic angular momentum (EAM) is conserved. The energy manifolds of donor and acceptor molecules can be engineered so that exciting the donor to a state with zero EAM precipitates a cutting event that results in a maximally-entangled pair of two acceptor molecules with opposite EAM polarity. Chains on both sides of the acceptor molecules produce Bell state wave packets that move apart ballistically. 

A quantum electrodynamics approach is used to elucidate coherent molecular QC, EAM conservation, and exciton entanglement. Spin-orbit coupling is assumed to be negligible, excitons are treated as two-level quasi-particles, and relaxation pathways that compete with QC are ignored for the sake of clarity~\cite{LaCount_2017}. A numerical implementation provides a proof-of-concept for the production of maximally entangled wave packets. Finally, we show that entangled states involving more than two EAM modes may also be created, providing a pathway to entangled exitonic qudits.

%

First consider a horizontal triad of acceptor-donor-acceptor arms in isolation, such as those connected by one of the black dashed lines in Figure \ref{Schematic}. This set has the potential to undergo spontaneous down-conversion, a three-body interaction in which the cooperative decay of an exciton on the donor arm results in the creation of excitons on each of the acceptor arms. In the initial state, $\ket{I_{\rm triad}}$,  the donor is in its first excited state and both acceptors are in their ground states. Cutting results in a final state, $\ket{F_{\rm triad}}$, for which the donor is in its ground state and both acceptors are in their first excited state. The complex-valued transition matrix element, $M = \braket{F_{\rm triad}|\hat{H}_{\rm triad}|I_{\rm triad}}$, controls the cutting. It represents an accumulated effect of two time orderings and photon interactions involving all possible virtual states of the donor. A quantum electrodynamical expression for this three-body coupling was previously derived~\citep{Jenkins1999} and experimentally implemented~\cite{Weingarten_2017} to demonstrate three-body upconversion dynamics.

Down-conversion can also occur among three-arm centrosymmetric molecules, where coupling between the arms on a given molecule causes level splittings in the eigenstates. The lowest-energy portion of this manifold can be obtained by diagonalizing a Hamiltonian based on the ground and first excited states of each arm:
\begin{equation}
\hat{H}  = \sum_{n = 0}^2 \hat{H}_0^n  + \hat{H}_{\mbox{\tiny QC}}
\label{H1}
\end{equation}
where
\begin{eqnarray}
\hat{H}_0^n &=&  \sum_{j=1}^3\Delta_n \hat{c}_{j}^{n\dagger}\hat{c}_{j}^n  + \sum_{\braket{ j,j' }} \tau_n \hat{c}^{n\dagger}_{j'}\hat{c}_{j}^n  + H.c.\nonumber \\
\hat{H}_{\mbox{\tiny QC}}  &=& \sum_{j=1}^3 M \hat{c}_{j}^{1\dagger}\hat{c}_{j}^{2\dagger} \hat{c}_{j}^0 + H.c.
\label{H2}
\end{eqnarray}

Here $\hat{c}_j^{n\dagger}$ is the creation operator for the first excited state on arm $j$ of molecule $n$, with the donor denoted by n = 0 and identical acceptors by n = 1 and 2. The operators obey bosonic commutation relations $[\hat{c}_{j}^n, \hat{c}_{j}^{n\dagger}]_-=\delta_{ij}$. Summation over nearest neighbors is indicated with $\braket{ j,j' }$. The excited state energy of each arm is $\Delta_n$, and arm-to-arm coupling, $\tau_n$, is approximated as being due to dipole-dipole interactions between nearest neighbors. Quantum electrodynamics, with an electric dipole approximation, can once again be used to derive these complex-valued intramolecular hopping terms~\cite{Daniels_2003}. The acceptors are identical, so $\Delta_1=\Delta_2$, $\tau_1=\tau_2$, and only the first acceptor subscript is used. The arm-to-arm QC matrix element, $M$, implicitly accounts for the effect of the manifold of donor states without needing to explicitly identify them.  


It is straightforward to show~\cite{AndrewsPRL2013} that the ground state of each molecule is $\ket{0^n}=\prod_{j=1}^{3}\ket{\xi_{j,0}^n}$, where arm $j$ supports two energy levels: ground state $\ket{\xi_{j,0}}_n$ and excited state $\ket{\xi_{j,1}}_n$. The three lowest excited states of a given molecule are then
\begin{equation}
\ket{q}_n = \sum_{j=1}^{3}\frac{\varepsilon^{(j-1)q}}{\sqrt{3}}\ket{e_j}_n.
\label{TBstates}
\end{equation}

Phase shifts are in increments of $\varepsilon=\mathrm{e}^{\imath 2\pi/3}$, and arm occupations are described by
\begin{equation}
\ket{e_j}_n = \ket{\xi_{j,1}}_n\prod_{m\neq j}^{3}\ket{\xi_{m,0}}_n.
\end{equation}

The EAM, $q$, is an integer that can take on values of -1, 0, and +1. It has corresponding energies of~\cite{AndrewsPRL2013} 
\begin{equation}
E_{q}^n=\Delta_n + 2|\tau_n| \cos\biggl(\frac{2\pi q}{3}\biggr).
\label{TBenergies}
\end{equation}

\begin{figure}[hptb]\begin{center}
\includegraphics[width=0.5\textwidth]{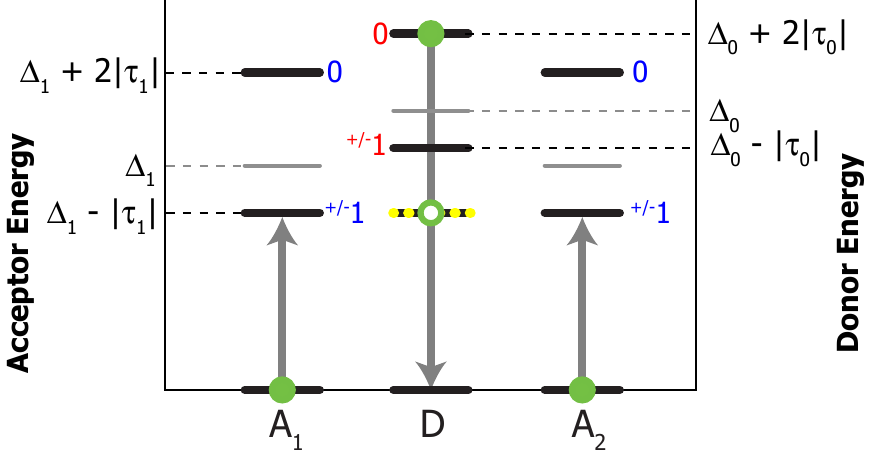}
\caption{\emph{Energy diagram for cooperative quantum cutting}. Centrosymmetric, three-arm donor and acceptor molecules have eigenstate energies shown by solid black lines. The electronic energy of virtual donor states is indicated with alternating yellow and black. Each excited state is labeled with an EAM, blue for acceptors and red for the donor. Excitation of the zero EAM (highest energy) state of the donor can cause a three-body interaction in which the acceptors are each excited to states that have an EAM of magnitude one.}
\label{Cooperative_Cutting}
\end{center}
\end{figure}

%
These energies decrease with increasing EAM and are degenerate with respect to its sign. Laser excitation of the zero EAM donor state can precipitate a three-body interaction in which the acceptors are each excited to states that have an EAM of magnitude one. This is shown in Figure \ref{Cooperative_Cutting}, where it is clear that resonant cutting occurs when $E_0^0 = 2 E_1^1$---i.e. when
\begin{equation}
\Delta_0 + 2|\tau_0| =  2 (\Delta_1 -  |\tau_1|) .
\label{Resonance}
\end{equation}
%

%

It is therefore possible to use energy down-conversion to produce EAM in the acceptor molecules, although energy considerations are mute as to the direction of their rotation. Symmetry does constrain these directions, though, and a simple selection rule can be derived. First, use Equation \ref{TBstates} to write out the initial and final states as
\begin{eqnarray}
\ket{I} &=& \sum_{r = 1}^3 \frac{1}{\sqrt{3}} \ket{e_r}_0 \ket{g}_1 \ket{g}_2 \nonumber \\
\ket{F} &=& \sum_{s, t = 1}^3 \frac{1}{3} \varepsilon^{(s-1)q_1} \varepsilon^{(t-1)q_2}  \ket{g}_0 \ket{e_s}_1 \ket{e_t}_2 ,
\label{InitFinal}
\end{eqnarray}
where $\ket{g}_n$ represents ground states. The acceptor excitons are assumed to have EAM of $q_1$ and $q_2$ but are otherwise unconstrained. Combining Equations \ref{H1} and \ref{InitFinal} then gives
\begin{equation}
\braket{F|\hat{H}|I} = \frac{M}{3^{3/2}} \sum_{j = 1}^3 \varepsilon^{(j-1)(q_1 + q_2)} .
\end{equation} 
However, cyclic sums of periodic exponentials have a simple orthogonality property:
\begin{equation}
\sum_{j = 1}^3 \varepsilon^{(j-1)(m-n)} = 3 \delta_{m,n},
\end{equation}
where $\delta_{m,n}$ is the Kronecker delta function. This implies that 
\begin{equation}
\braket{F|\hat{H}|I} = \frac{M}{\sqrt{3}} \delta_{q_1, \null^-q_2} .
\label{selection_rule}
\end{equation} 

The EAM of the acceptor molecules must therefore be of opposite sign in any product state, and EAM is conserved in the molecular downconversion. 

For the energy alignment of Figure \ref{Cooperative_Cutting}, Equations \ref{InitFinal} and \ref{selection_rule} imply that the system actually evolves as a linear combination of just three states:
\begin{eqnarray}
\ket{I} &:=& \ket{0}_0 \ket{g}_1 \ket{g}_2 \equiv \ket{1,0,0} \nonumber \\
\ket{F_1} &:=& \ket{g}_0 \ket{\null^+1}_1 \ket{\null^-1}_2 \equiv \ket{0, \null^+1, \null^-1}\nonumber \\
\ket{F_2} &:=& \ket{g}_0 \ket{\null^-1}_1 \ket{\null^+1}_2 \equiv \ket{0, \null^-1, \null^+1}.
 \label{IandF} 
\end{eqnarray}
Here a condensed notation has been introduced at right. The degree of EAM entanglement depends on the relative weighting of acceptor states $\ket{F_1}$ and $\ket{F_2}$ over time. Since the initial condition is symmetric with respect to their population, and their rates of change are governed by equal matrix elements, $\braket{F_1|H|I} = \braket{F_2|H|I}$, the acceptor states must always have the same amplitude. The time-varying state vector is then
\begin{equation}
\ket{\Psi(t)} = \frac{u_a(t)}{\sqrt{2}} \bigl( \ket{0,\null^+1, \null^-1} + \ket{0,\null^-1, \null^+1} \bigr) + u_d(t) \ket{1,0,0} .
\label{Bell2}
\end{equation}
 
Down-conversion therefore results in the production of an excitonic Bell state in which opposite-polarity EAM states are maximally entangled. The population of this state varies cyclically but the degree of entanglement between the angular momentum polarities is fixed. 

The time scale for population transfer is obtained by constructing a Hamiltonian matrix, from Equation \ref{H1}, that operates on the coefficients of Equation \ref{Bell2}:
\begin{equation}
[H] = 
 \begin{bmatrix}
  2E_1^1 &  M\sqrt{\frac{2}{3}} \,  \\
   & \\
 M^*\sqrt{\frac{2}{3}}  & \gamma 2E_1^1 \\
 \end{bmatrix} .
 \label{Hmx1}
\end{equation}
Here a dimensionless detuning factor, $\gamma$, pre-multiplies the donor energy so that values other than unity represent non-resonant conditions. The frequency of population transfer due to cutting, $\omega_{\mbox{\tiny QC}}$, is then found to be
\begin{equation}
\omega_{\mbox{\tiny QC}} = 2\sqrt{\frac{2 |M|^2 +  3(\gamma -1)^2(E_1^1)^2}{3}}.
\end{equation}

Detuning ($\gamma\neq 1$) increases the oscillation rate but does not change the degree of entanglement between the two acceptor states.  However, expressions for amplitudes $u_d(t)$ and $u_a(t)$ are easily obtained, and these can be used to quantify the reduced von Neumann entropy associated with a single acceptor state. Tracing over the first and third dimensions of the state operator gives the following reduced version:
\begin{equation}
\hat\rho_{1}= \frac{|u_a|^2}{2} (\ket{\null^+1}\!{}_{1}\bra{\null^+1}\!{}_{1} +\ket{\null^-1}\!{}_{1}\bra{\null^-1}\!{}_{1})+   |u_d|^2\ket{0}\!{}_{1}\bra{0}\!{}_{1}
\end{equation}

\begin{figure}[hptb]\begin{center}
\includegraphics[width=0.45\textwidth]{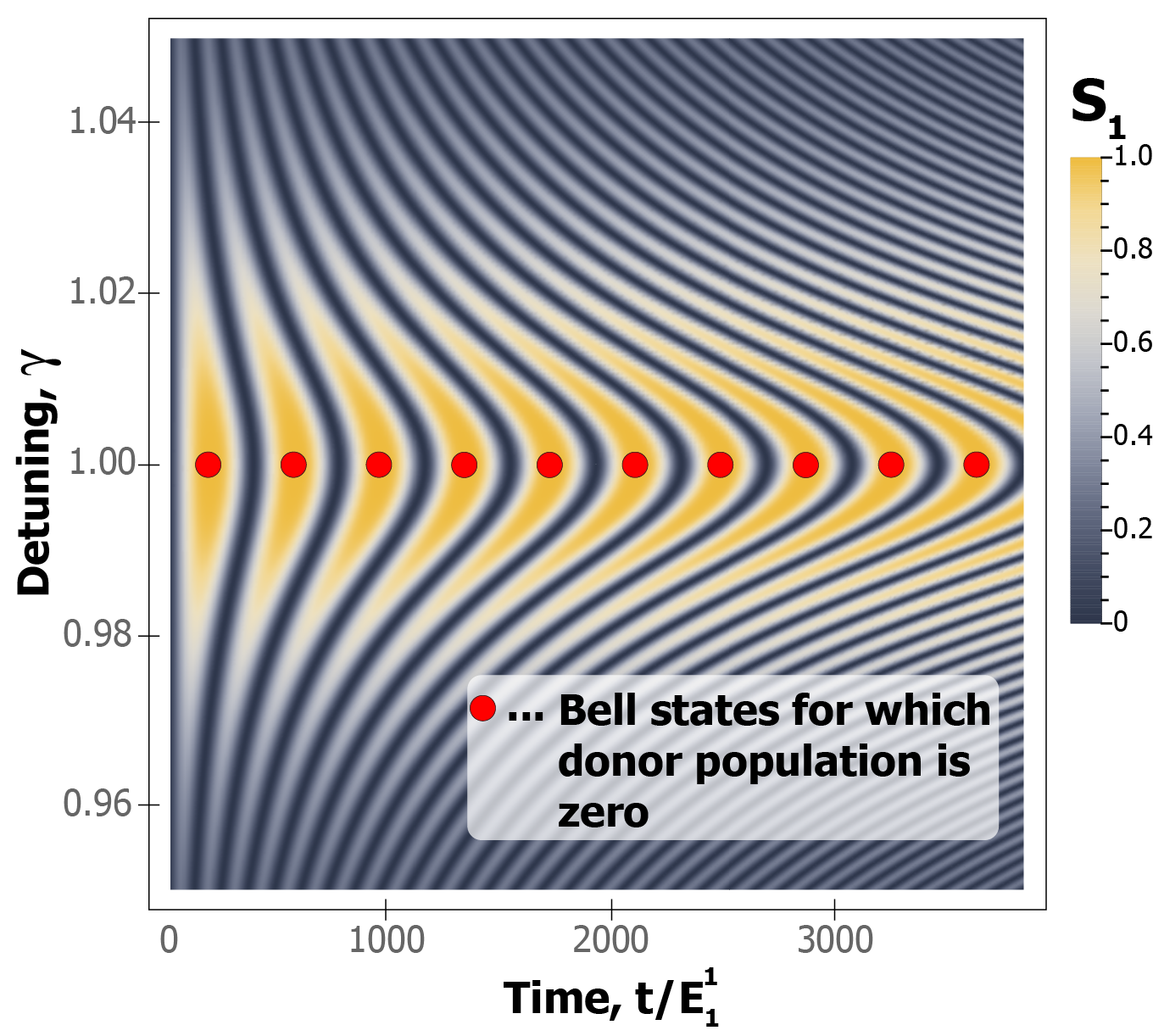}
\caption{\emph{Angular momentum entanglement}. The reduced von Neumann entropy of a single acceptor, $S_1$, is plotted as a function of time and the detuning, $\gamma$. This quantifies the entanglement of one acceptor with respect to the entire system, but the Bell state shared by the two acceptors is time invariant.}
\label{Entropy}
\end{center}
\end{figure}

%
The reduced von Neumann entropy, $S_1 = -\hat\rho_1 \cdot \log_2 (\hat\rho_1)$, can then be quantified. This measure of the entanglement of one acceptor with the rest of the system is plotted in Figure \ref{Entropy}.

%
%
%

No assumption has been made other than to disregard the possibility of exciting a zero EAM state in each acceptor (Figure \ref{Cooperative_Cutting}). This is reasonable provided that the QC coupling, $M$, is sufficiently small relative to the hopping parameter, $\tau_1$, which controls the degree of acceptor energy splitting. It is worth noting that these time-invariant acceptor Bell states are distinct from those associated with the mid-point of generic resonant energy transfers which have been elicited in quantum dot dimers~\cite{Quiroga_1999}. 

\begin{figure}[hptb]\begin{center}
\includegraphics[width=0.5\textwidth]{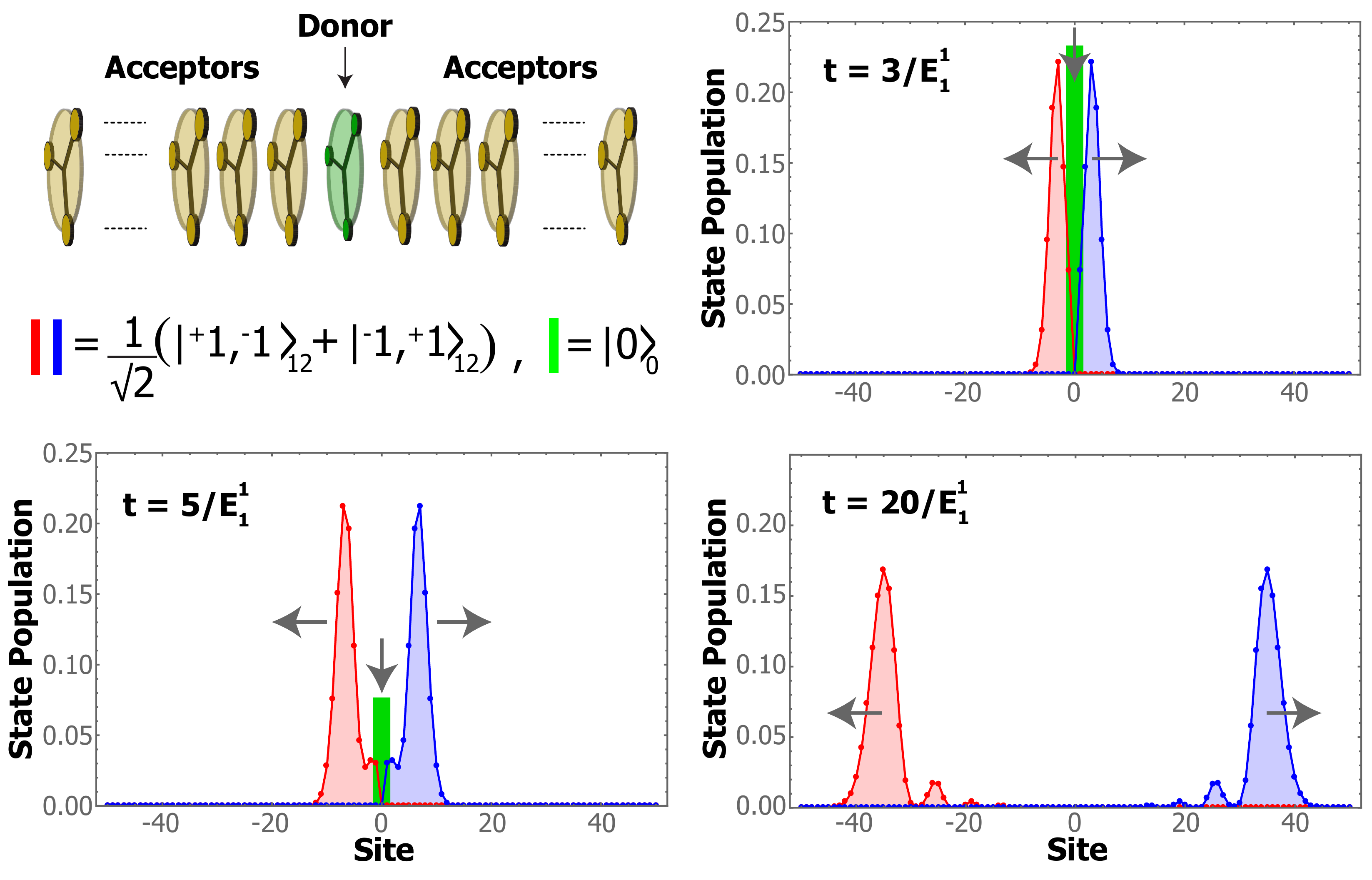}
\caption{\emph{Entangled exciton wave packets}. A donor molecule (site 0) is initially excited into a state with no EAM (green). Resonant QC to its nearest neighbors produces a pair of excitons in a Bell state with EAM of $\pm 1$. Entangled wave packets are formed which move in opposite directions. Parameters: $E_1^1 = \eta/2$, $M = \eta/6$. Red and blue colors highlight the left and right components of the same, two-particle excitonic state.}
\label{Entangled_Packets}
\end{center}
\end{figure}

A direct excitonic counterpart to the spontaneous parametric downconversion of photons can now be constructed by adding a chain of cofacial acceptor molecules to either side of the assembly as shown in the upper-left panel of Figure \ref{Entangled_Packets}. Let  $\hat{b}_n^{\dagger}$ be the creation operator for an excitonic Bell state on molecular sites $-n$ and $+n$, with  $[\hat{b}_{n}, \hat{b}_{m}^{\dagger}]_-=\delta_{mn}$.  The donor, at site zero, has its own operator for the creation of a state with zero EAM, $\hat{b}_0^{\dagger}$. The dynamics are then captured by the following Hamiltonian:
\begin{eqnarray}
\hat{H}_{\rm chain} &=&  \sum_{n=1}^L 2 E_1^1 \hat{b}_{n}^{\dagger}\hat{b}_{n}  + \sum_{\braket{m,n}=1}^L \eta \hat{b}_{m}^{\dagger}\hat{b}_{n} \nonumber \\
        & & + E_0^0 \hat{b}_{0}^{\dagger}\hat{b}_{0}  + 3 M \hat{b}_{\null^{-}1}^{\dagger}\hat{b}_{\null^{+}1}^{\dagger} \hat{b}_{0}   + H.c.
\label{Hchain}
\end{eqnarray}
%


Here $\eta$ is the coupling between states associated with adjacent acceptors, and there are a total of $2L+1$ molecular sites. Excitation of the donor produces entangled, twisted exciton wave packets that travel in opposite directions. This is shown in Figure \ref{Entangled_Packets} for resonant down-conversion.

%
%
%

While three-arm molecules were used for the sake of clarity, the dynamics are richer for $N$-arm molecules with $N > 4$. These support integer EAM bounded by $-(N-1)/2$ and $(N-1)/2$, allowing for the production of excitonic Bell states with a range of excitonic angular momenta. It also makes possible the off-resonant excitation of entangled states involving multiple EAM, allowing for the construction of entangled qudit states. For instance, a five-arm assembly can be used to generate product states of the form
\begin{eqnarray}
\ket{F} &=&  \frac{c_1}{\sqrt{4}} \ket{g}_0 \bigl(\ket{\null^+1, \null^-1}_{12} + \ket{\null^-1, \null^+1}_{12} \bigr)  \nonumber \\
& &+  \frac{c_1}{\sqrt{4}}\ket{g}_0 \bigl(\ket{\null^+2, \null^-2}_{12} + \ket{\null^-2, \null^+2}_{12}\bigr)  \\
& &  + c_2 \ket{g}_0 \ket{0,0}_{12} . \nonumber
\end{eqnarray}
\label{4-state}

\begin{figure}[hptb]\begin{center}
\includegraphics[width=0.4\textwidth]{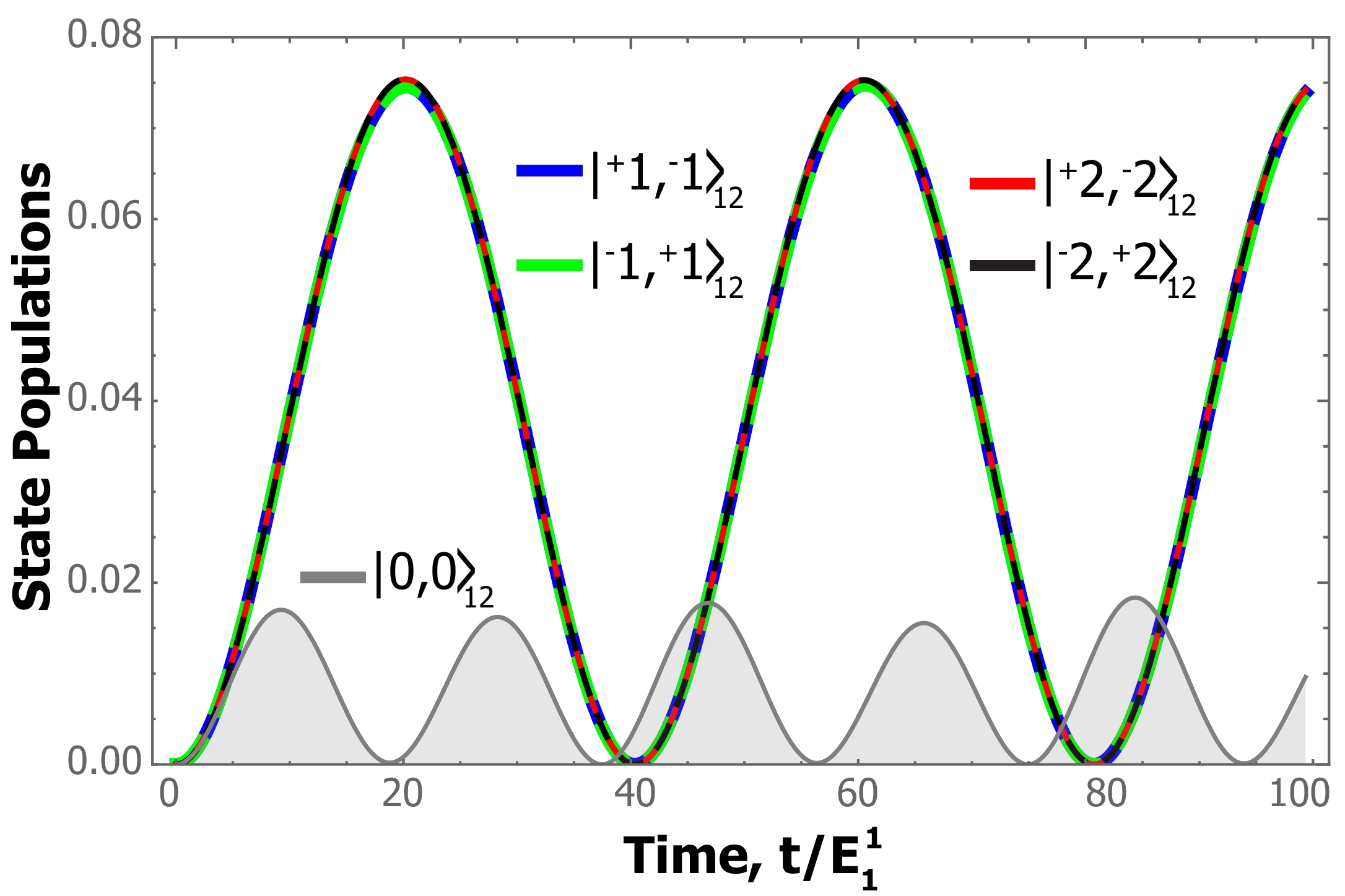}
\caption{\emph{Five-state entanglement}. The donor state of an assembly of three five-arm molecules is initially excited into a state of with no EAM. The quantum cutting that ensues produces an acceptor state in which 4 EAM ($\pm 1$, $\pm 2$) are entangled. A small acceptor component with no EAM (gray) is also generated for the parameters chosen: $M = 0.05 E_1^1$, $\tau_1 = \Delta_1/15$, $\gamma = 1.077$.}
\label{4-State_Entanglement}
\end{center}
\end{figure}

This is numerically demonstrated in Figure \ref{4-State_Entanglement} by constructing a 26-dimensional ($N^2+1$) Hamiltonian matrix and solving for the evolving coefficients of all possible combinations of acceptor EAM. The product states other than those shown remain unpopulated, consistent with the EAM selection rule of Equation \ref{selection_rule}.

%
%
%

In conclusion, it is possible to use quantum cutting to produce exciton pairs with maximally entangled EAM of opposite polarity. A chain of sites can be added to either side of the assembly to produce entangled, twisted exciton wave packets amenable to spin-chain protocols for entanglement transfers ~\cite{Bose_2003}. Systems composed of molecules with more than four arms allow for Bell states with EAM of magnitude greater than one as well as entangled states involving more than two EAM. Centrosymmetric scaffold structures may be useful in arranging chromophores with the requisite orientation and spacing. Splitting of the excitonic energy levels on each molecule is controlled by the electronic interaction between the adjacent arms, and making this strong improves cutting selectivity. Competitive relaxation pathways, such as resonant energy transfer, internal conversion, and photoluminescence can be managed by choosing molecules with proper energy level alignment~\cite{LaCount_2017}. As in optical settings, the ability to produce entangled states is an enabler for foundational experiments and technologies based on non-local quantum correlations. 


\end{document}